\documentclass[aip,amsmath,amssymb,reprint]{revtex4-1}
\usepackage{graphicx}
\usepackage{dcolumn}
\usepackage{bm}
\usepackage[caption=false]{subfig}

\usepackage[utf8]{inputenc}
\usepackage[T1]{fontenc}
\usepackage{mathptmx}
\usepackage{etoolbox}
\usepackage{braket}
\usepackage{multirow}
\usepackage{svg}
\usepackage{textcomp,gensymb}

\graphicspath{ {./images/} }

\begin{document}

\preprint{AIP/123-QED}

\title{High quality entanglement distribution through telecommunication fiber using near-infrared non-degenerate photon pairs}

\author{Rui Wang}
 %\altaffiliation[Electronic mail : ]{E0305650@u.nus.edu}%
\author{Anindya Banerji}%
\altaffiliation[Electronic mail: ]{cqtab@nus.edu.sg}%
\affiliation{ 
Centre for Quantum Technologies, National University of Singapore, 3 Science Drive 2, S117543, Singapore}%

\author{Alexander Ling}%
\affiliation{ 
Centre for Quantum Technologies, National University of Singapore, 3 Science Drive 2, S117543, Singapore}%
\affiliation{ 
Physics Department, National University of Singapore, 2 Science Drive 3 ,S117542, Singapore}%

\date{\today}

\begin{abstract}
For practical quantum communications, the efficiency of the entire system (source, quantum channel and detectors) must be taken into account. In many urban environments, the quantum channel in the form of telecommunication optical fiber (confirming to ITU G.652D standards) are available, but the detectors in this range typically have low efficiency. We investigate the possibility that for campus-type communications, entangled photons prepared in the Near-Infrared Range (NIR) can be transmitted successfully while preserving polarization entanglement. We demonstrate the distribution of degenerate and non-degenerate entangled photon pairs of wavelength around 810 nm through standard telecommunication fiber. This technique benefits from the high efficiency of the NIR single photon detectors and the mature design of setups around 810 nm.. In this work, we obtain high quality entanglement (visibility is 94.8\% based on the raw data) after an overall distance of 12 km, corresponding to about -36 dB of fiber induced loss.
\end{abstract}

\maketitle

\section{Introduction}
Quantum entanglement has emerged as the pivotal concept in a variety of applications, such as quantum teleportation\cite{Bouwmeester1997}, quantum key distribution (QKD)\cite{Gisin2002}, quantum metrology\cite{Fink2019}, distributed quantum sensing and quantum computation. It is also going to play an important role in the emerging field of quantum networks which would involve entanglement distribution across the network and the teleportation of quantum states between nodes\cite{Gisin2007,Kimble2008}. In all these applications of entanglement, the quality of entanglement being distributed over a network will have a significant impact.
\par Entangled photons generated from spontaneous parametric down conversion processes (SPDC) \cite{Kwiat1995,Shalm2015,Giustina2015} have been the workhorse for experiments in quantum information. There have been several demonstrations of quantum communication protocols using these sources over free-space links \cite{Kurtsiefer2002} as well as dedicated fiber links \cite{Poppe2004}. These experiments benefit from the combined high quality performance of the source in the near infrared regime of around 800 nm as well as the easy availability of compact and efficient semiconductor detectors based on silicon Geiger- mode avalanche photodiodes (Si GM-APD) with excellent performance.
\par Interestingly, the choice of wavelength poses an obstacle when integrating these sources with existing optical communication infrastructure that is designed to benefit from the low-loss transmission window around 1550 nm. Potential solutions could involve frequency conversion of the single photons to match the transmission window in case of the existing sources or the development of high performance entangled photon sources that operate in the longer wavelength regime. However, existing semiconductor detectors like InGaAs GM-APD designed for these wavelengths suffer from low efficiencies, high dark count rates and longer dead times. These factors negatively affect the overall system performance unless using superconducting single photon detectors and accepting their corresponding power requirements.
\par It might be possible to perform entanglement distribution with NIR signals when the use-case is for relatively short distances, e.g. within a campus or data-centre. This is due to the fact that although the shorter wavelengths experience higher levels of attenuation in these fibers (-3 dB/km compared to -0.22 dB/km), the greater efficiency of the Si GM-APDs can be used to offset the effect of fiber attenuation for short distances. In fact, it has been shown\cite{Meyer2010,Holloway2011} that this translates to a lower system loss suffered by 800 nm photons compared to 1550 nm photons over a transmission distance of about 2.4 km.
\begin{figure*}[htbp]
    \includegraphics[scale=0.45]{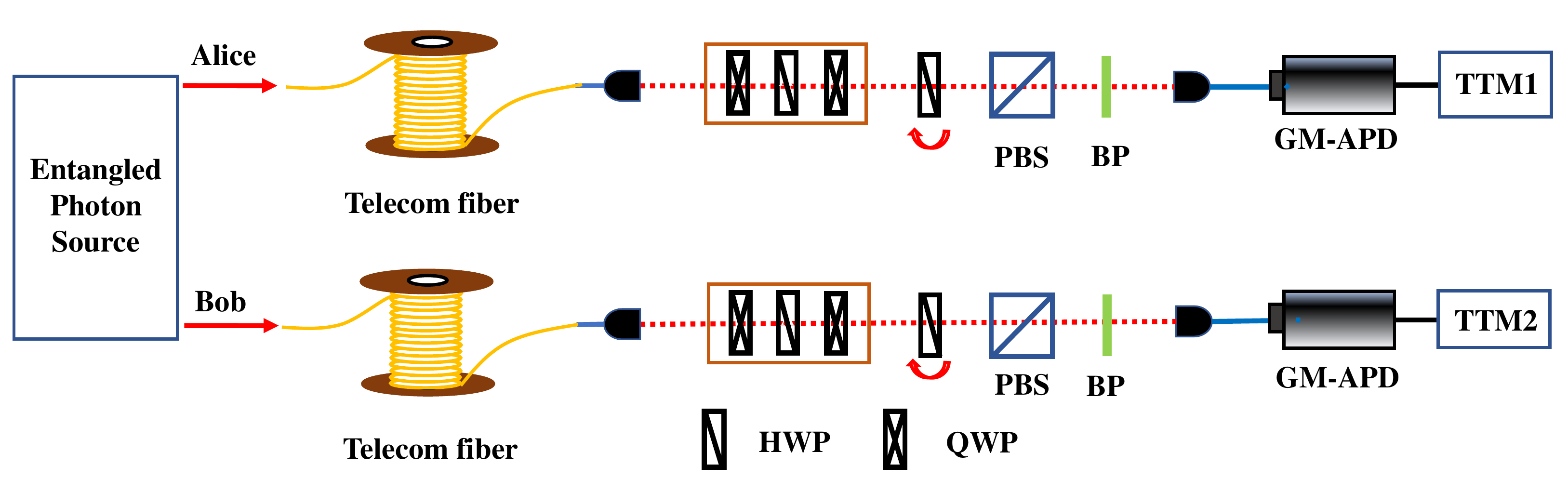}
    \caption{Schematic of the experiment. A PPKTP crystal is used to produce photon pairs at degenerate or non-degenerate wavelengths by controlling the crystal temperature. A WDM is used to separate the photon pairs into two arms. The separated photons in each arm are transmitted through standard telecom fiber before being detected by Si GM-APD. L1,L2: Aspherical lens, HWP : Half-wave plate, WDM : Wavelength Division Multiplexer, PBS : Polarization Beam Splitter, BP : Bandpass Filter. GM-APD : Geiger-mode Avalanche photodiode detector. Alice and Bob perform polarization state measurement on their photon using a half wave plate and polarization beamsplitter (PBS) whose output modes are directed to the GM-APD. Bandpass filters are used to further improve wavelength filtering. Detection events are recorded using a time-tagging module (TTM). From these tags, the cross correlation can be determined from delay histograms between the channels as well as the visibility can be calculated. }. 
    \label{fig:Exp_Setup}
\end{figure*}
\par The use of NIR wavelengths in fibers optimised for telecom wavelength introduces additional complication. These fibers are no longer single-mode for the shorter wavelengths. This leads to excitation of higher order modes giving rise to mode dispersion effects. This can affect the quality of the quantum signal resulting in reduced entanglement visibility. It was shown that the entanglement visibility can be improved substantially by parallel application of spatial filtering and precise temporal filtering to select only the fundamental mode. This resulted in an increase in the visibility by about 30\% but the coincidence counts reduced by almost a third \cite{Meyer2010}.
\par It should be noted that the above experiment used a degenerate SPDC process for creating entangled photons. This means that each photon of the photon-pair has identical wavelength. Degenerate phase matching conditions typically lead to a broader wavelength spectrum, which is this case was approximately 7 nm. On the other hand, SPDC processes can also be designed to operate in the non-degenerate condition. In this case, the signal and the idler photon has distinct wavelengths with a narrower FWHM linewidth of about 2 nm. It is interesting to study how the narrower linewidth entangled photons from non-degenerate SPDC process affects the excitation of higher order modes which in turn would affect the entanglement visibility. This is the scenario we investigate in this article.
\par This work demonstrates that polarization entanglement could be robustly distributed through telecommunication fiber using near-infrared non-degenerate entangled photons. We show that, only with temporal filtering applied, high quality entanglement after a transmission distance of 12 km, corresponding to fiber losses of -36 dB, could be obtained. This highlights an advantage that non degenerate entangled photon pairs hold over degenerate entangled photon pairs when it comes to distribution of quantum information in point-to-point communication over short distances through existing fiber infrastructure.
\par The rest of the article is arranged as follows. We first present the experimental setup and discuss the various implementation scenarios in section II. A symmetric distribution scenario is considered, which means that the receivers are equidistant from the source and are connected by similar lengths of telecommunication fiber. Section III consists of our results related to the quality of entanglement under different lengths of fiber connecting the source with the receivers. We conclude the article in section IV where we highlight some of our results and discuss future extensions.

\section{EXPERIMENT}
The experimental setup is shown in Fig.1. A type-0 periodically poled potassium titanyl phosphate (PPKTP) crystal under collinear phase-matching condition is used for generating polarization-entangled photon pairs\cite{Lohrmann2020}. After the crystal, the photon pairs are prepared in the following Bell state:
\[\Ket{\Phi^+} = \frac{1}{\sqrt2} (\Ket{H_A H_B} \pm e^{i\phi}\Ket{V_A V_B})\]
A wavelength division multiplexer (WDM) is used to separate the signal and idler photons into two arms. Then, we send the photon pairs through equal lengths of single mode telecom fibers (SMF28-e, ITU G652.D; mode field diameter MFD of 9.2 $\mu$m) in each arm. Polarization compensation at the end of the transmission channel is carried out by a combination of two quarter-wave plates and a half-wave plate. By using another half-wave plate and polarization beam splitter (PBS), we could analyse the correlation of entangled photon-pairs between two arms. Additional bandpass filters were used to improve wavelength filtering due to the imperfect WDM which only offered 15 dB isolation. Finally, the photon is detected by Si GM-APD. The state and the time of detection was recorded by a time-tagging module and the timestamp information is analysed for the cross-correlation between the two arms.

\section{RESULTS}
In a quasi phase-matched crystal, the spectrum of the emitted pair of photons can be controlled via temperature tuning of the crystal. We study this distribution in Fig. 2 measured with a spectrometer. Degenerate phase matching is achieved for a crystal temperature of 26.5 $^{\circ}$C. At this temperature, the spectrum of the photons for both the arms are centred around 810 nm with a bandwidth of around 7nm. Increasing the temperature of the crystal to 34 $^{\circ}$C shifts the wavelength of the signal photon to 774 nm and that of the idler photon to 850 nm. This corresponds to the non-degenerate phase matching condition. The bandwidth of the signal and idler photons is much narrower compared to before and is recorded to be around 2 nm.

\begin{figure}[h]
    \includegraphics[width = 0.95\linewidth]{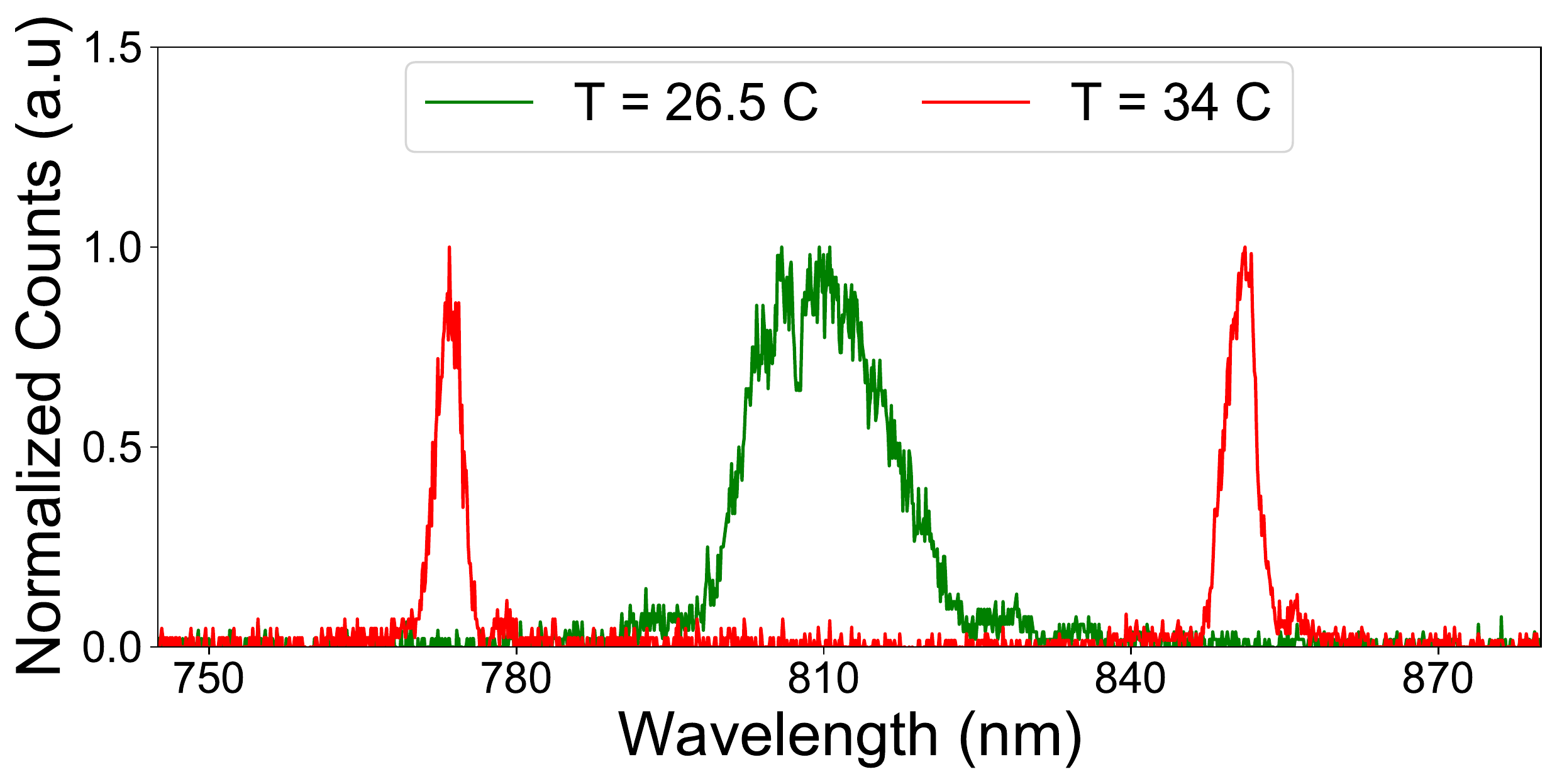}
    \label{fig:Spectrum}
    \caption{Normalized spectrum of entangled photon source. The spectrum of the non-degenerate entangled photons (red line) has a linewidth (FWHM) of about 2 nm and that of the degenerate entangled photons (green line) has a linewidth of about 7 nm. }
\end{figure}

The cross-correlation measurement is performed for both degenerate and non-degenerate photon pairs. As the fiber we use conforms to ITU G.652D standard \cite{ITU-T2016}, it behaves as a few-mode fiber for near-infrared photons\cite{Gloge1971}. The total number of modes that can exist in the fiber for such wavelengths can be obtained by calculating the normalized frequency $V_{norm}$ based on the datasheet of the fiber we used. The calculated $V_{norm}$ number is 4.3. This means multiple modes could exist inside the fiber. According to guided wave theory\cite{Paul1996}, there should be two linearly polarized modes of propagation, $LP_{01}$ and $LP_{11}$. Therefore, if we perform a cross-correlation measurement from the timing data of the two channels, multiple peaks should appear beside a central peak since the group velocity of higher order modes differ from that of the fundamental mode. We show the results of the cross correlation measurement in Fig. 3. 
\begin{figure}[h]
\includegraphics[width = 0.95\linewidth]{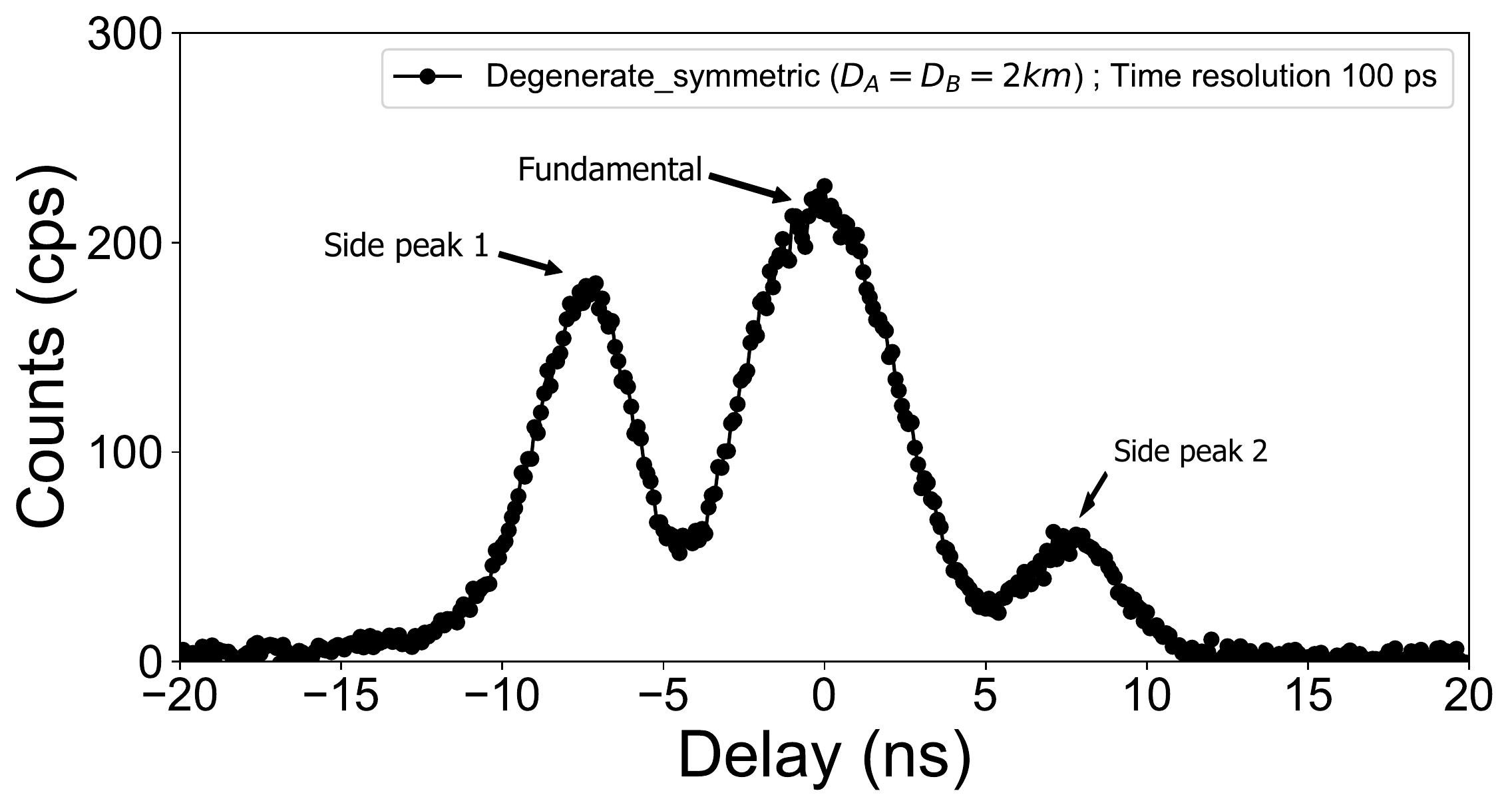} 
\includegraphics[width =0.95\linewidth]{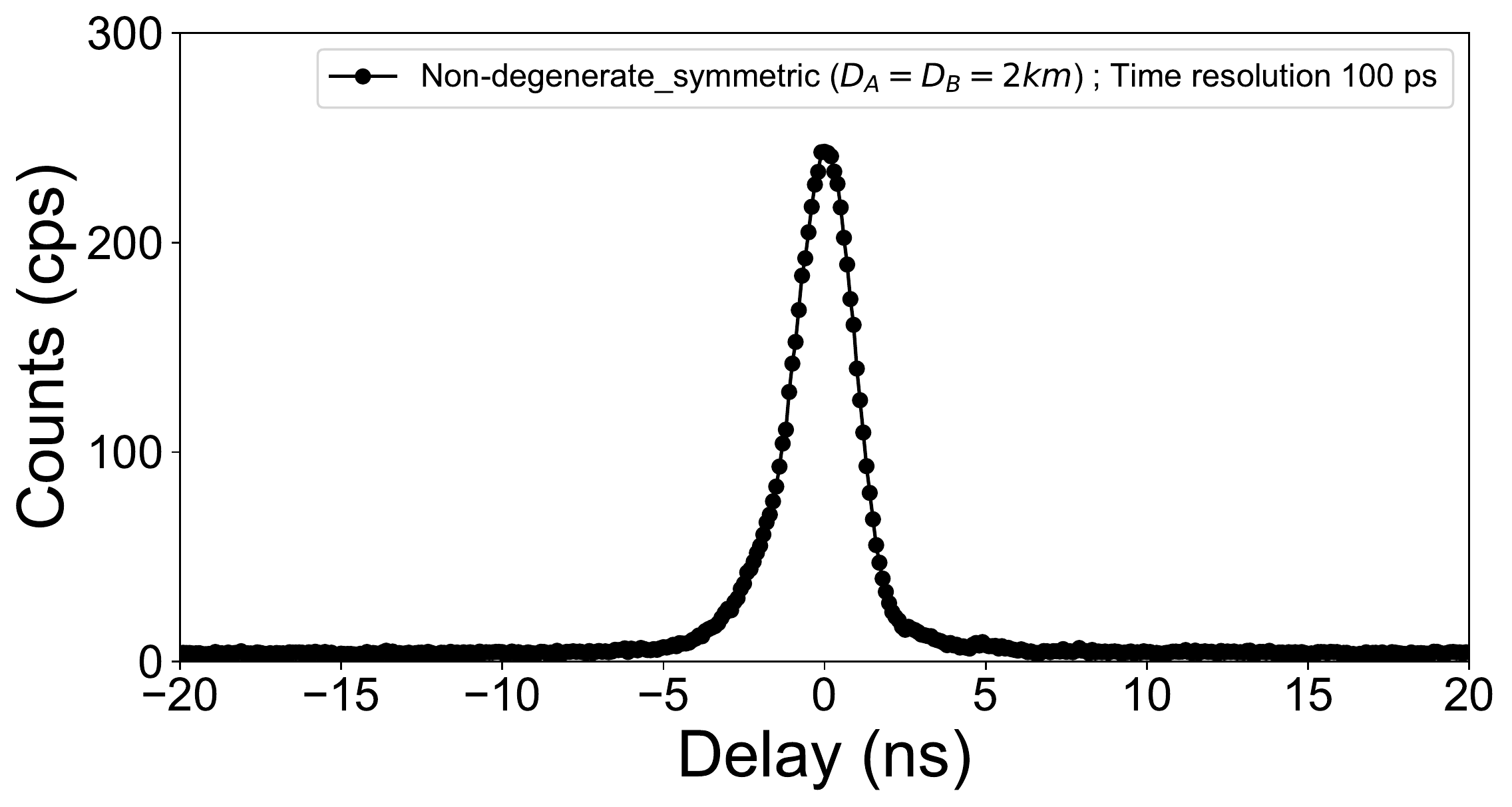}
\caption{The cross-correlation measurement at a symmetric situation {(equal length of fiber in each arm)}. (a) Degenerate situation . The distance between the fundamental mode and higher order mode is proportional to the fiber length. (b) Non-degenerate situation. The side peaks are absent.}
\label{fig:Cross_corr}
\end{figure}
In the degenerate case, as in Fig. 3(a), a central peak is obtained from the photons that arrive at the detector at the same time. Side peak 1 on the left of the fundamental is due to the coincidences between the $LP_{01}$ mode in Alice's arm and $LP_{11}$ mode in Bob's arm. Side peak 2 on the right of the fundamental arise from coincidences between $LP_{11}$ mode is Alice's arm and $LP_{01}$ mode in Bob's arm. The separation between the fundamental and the side peaks increases with increasing fiber length. However, as we move from degenerate to non-degenerate condition by increasing the crystal temperature, we see that the side peaks disappear and only a central peak is left as shown in Fig. 3(b). This indicates that only the fundamental mode is strongly excited in the fiber. This is due to the reduced bandwidth of the entangled photons in the non-degenerate situation.\par
The excitation of higher order modes leads to a reduced number of photons in the fundamental modes which affects the detection rates adversely. This suggests that the pair detection rates for non-degenerate entangled photons prepared in NIR would be higher compared to degenerate photon pairs. This could result in a longer effective transmission distance over which NIR entangled photons outperform those prepared in telecom regime.\par
In order to quantify this comparison, we could assume the same structure of the telecom photon source with our case (similar source brightness), and a similar transmission curve for the WDM. Additionally, we assume Si GM-APD efficiency of about 50\% for the NIR signal and commercial state of the art InGaAs GM-APD efficiency of 25\% \cite{id230} for the longer wavelength. The results of this comparison are shown in Fig. 4 where we study the variation in detected pair rates in the fundamental mode with fiber length. 
\begin{figure}[h]
    \includegraphics[width = 0.97\linewidth]{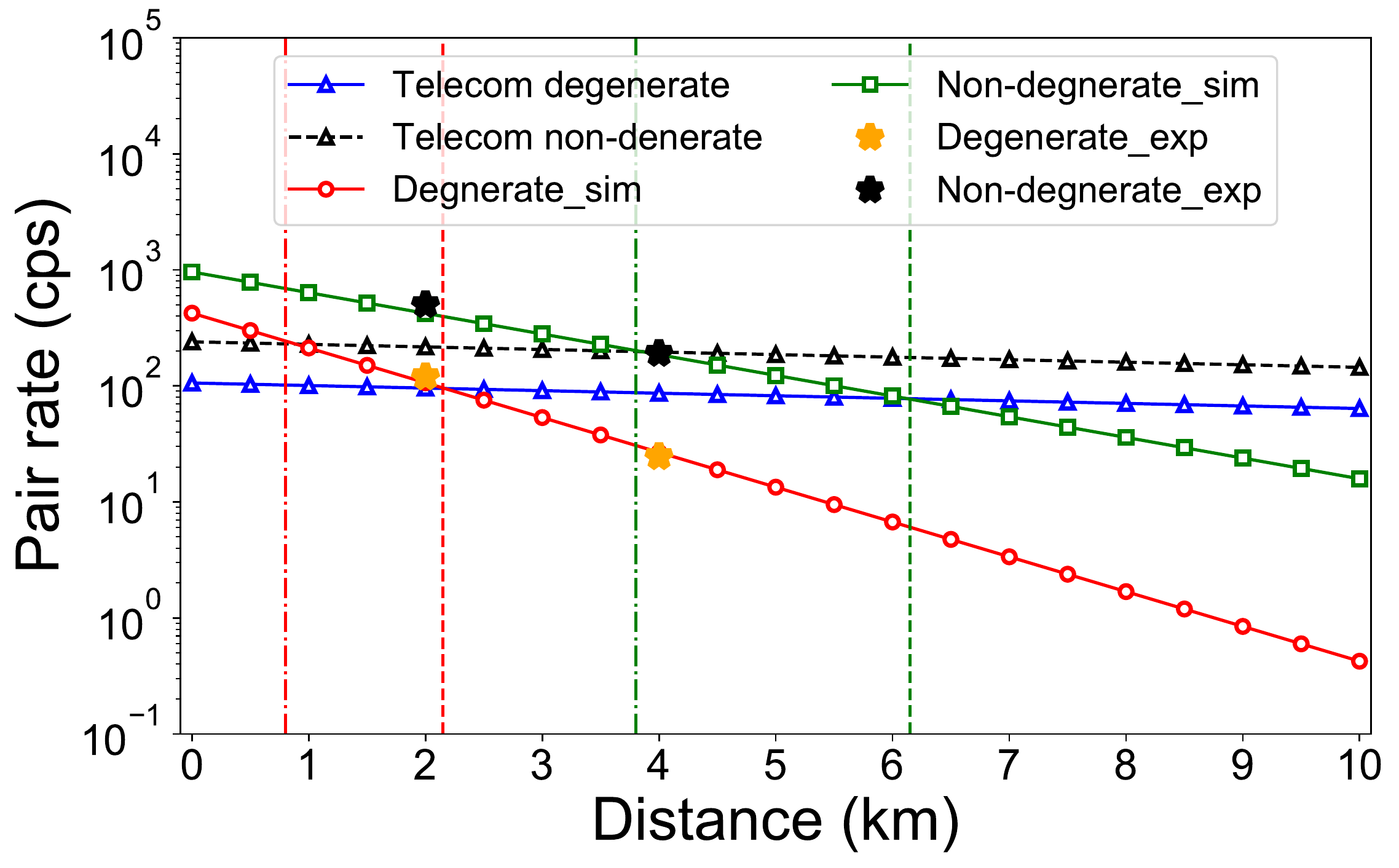}
    \caption{Comparison of the relationship between the detected pair rate and transmission distance with different types of entangled photon sources. Telecom degenerate (non-degenerate) refers to an entangled photon source operating in the degenerate (non-degenerate) situation with similar parameters as the NIR entangled photon source used in this work. The vertical green dot-dashed (dashed) line marks the distance at which the non-degenerate NIR system would no longer outperform a non-degenerate (degenerate) telecom system purely from a loss perspective. The red vertical lines mark the same for degenerate NIR entangled photons. The values corresponding to the black and yellow stars are obtained experimentally and are in excellent agreement with the predicted values.} 
    \label{fig:Comparison}
\end{figure}
In the model, we assume that the signal photon is detected at source while the idler photon is detected after being transmitted through the fiber. The blue solid line and the black dashed line is for degenerate and non-degenerate entangled photons prepared in the telecom wavelength. The intrinsic fiber loss is very low in this case. Since the fiber is single mode for this wavelength, no higher order modes are present. This results in a very gradual loss in the count rates. The red and green solid lines represent the situation for degenerate and non-degenerate entangled photon pairs prepared in NIR. Since this wavelength is not optimum for the fiber, the intrinsic fiber attenuation is very high. However, the absence of higher order modes leads to increased photon counts in the fundamental mode for the non-degenerate case leading to lower overall loss compared to the degenerate case. This increases the effective distance over which NIR entangled photon pairs continue to outperform telecom wavelength entangled photon pairs from 2.4 km as previously reported to about 6 km. This means that non-degenerate NIR entangled photon systems compare favorably with a telecom system for up to -18 dB of fiber loss. We note that the photon pair count for the non-degenerate case is greater than the degenerate situation at the initial stage before the start of transmission due to the WDM. Under the degenerate situation, the WDM cannot split the photons efficiently between the two arms, behaving similar to a beam splitter. This reduces the coincidence counts between the two arms.\par
\begin{figure}[h]
    \includegraphics[width = 0.95\linewidth]{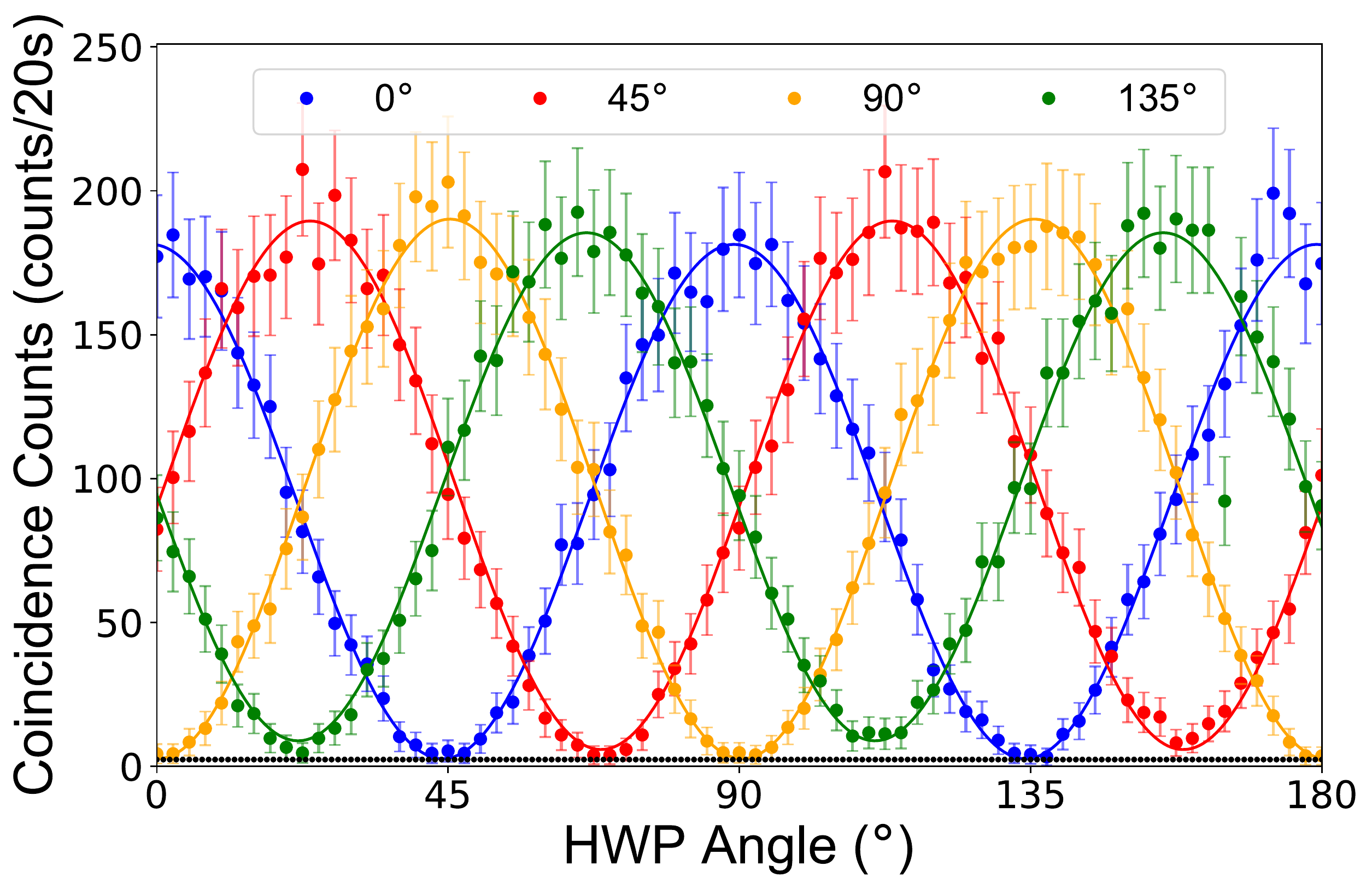}
    \caption{Polarization correlation in both H/V and D/A bases measured after 6 km telecom fiber in each arm corresponding to -36 dB of fiber loss and an estimated detector efficiency of 50\%. For each setting of the HWP in the signal arm, the HWP in the idler arm is rotated through 180$^{\circ}$.}
    \label{fig:Entanglement_visibility}
\end{figure}
In order to ensure high entanglement visibility, it is important to select only the fundamental mode in both arms for reasons mentioned earlier. The polarization compensator can then be tuned to offset the polarization rotation experienced by the fundamental mode. There are two strategies to elminate the higher order modes. One of them is to separate the two modes and compensate them individually, but it is difficult to extract all the higher order modes accurately. The other method is to use precise temporal and spatial filtering as mentioned before. Temporal filtering can be implemented by using narrow coincidence time windows while a spatial filter is implemented by coupling the photons to a short single mode fiber after transmission through the telecom fiber. The presence of the spatial filter increases the overall optical loss. However, as observed from the cross-correlation measurement, in the non-degenerate situation, the side peaks were absent and only the fundamental mode was present. This signifies that only a temporal filter should be sufficient to detect the actual coincidences.\par  
In order to verify the quality of the entanglement after transmission through several kilometers long telecom fiber by only applying temporal filter, we measured the entanglement visibility\cite{Hubel2007} and raw coincidences by using the coincidence window as a temporal filter. Following Fig. 4, we installed 6 km fiber spools in each arm of the experiment as in Fig. 1. This resulted in a separation of 12 km or a total of -36 dB of fiber loss between Alice and Bob. We used a combination of quarter and half waveplates as polarization compensator to compensate polarization perturbation introduced by fiber transmission. Two bandpass filters were used to clean the spectrum at the corresponding arm because of the imperfection of the WDM. Another half waveplate and a polarizing beam splitter were used in each arm to set the measurement bases. The pump power and the coincidence window were set at 0.5 mw and 1 ns respectively. The visibility measurement is shown in Fig. 5. In the H/V basis, the raw visibilities were 97.1\% and 97.2\% while in the D/A basis, they were 94.0\% and 90.8\%. This leads to an average visibility of 94.8\%, without any spatial filtering. For comparison, the average entanglement visibility was recorded to be 98.1\% at source and 96.1\% after 4 km separation. These results highlight that high quality entanglement can be preserved over an effective distance of 12 km using non-degenerate NIR photons through telecom fiber.

\section{Conclusion}
In this article, we have demonstrated the potential of distributing high quality entanglement through existing fiber infrastructure for point-to-point quantum communication. We have shown that using non-degenerate entangled photon pairs prepared in NIR, entanglement can be distributed over much longer distances compared to degenerate photon pairs for the same transmission and detection setup. This technique can prove advantageous owing to the large availability of entangled photon sources with high brightness in the NIR as well as compact and cost effective detection setups employing Si GM-APDs, and can find application in campus-wide implementations or short metropolitan networks.\par
Further, these results have implications for some implementations of quantum key distribution systems over short fiber links, for example, in-campus implementations. Our results show that entanglement based BBM92 protocol can be implemented in such situations as opposed to the more popular decoy state BB84 protocol. In a future iteration, we hope to build a fully functional QKD system using the technique developed here and test it over deployed fiber.

\begin{acknowledgments}
This research is supported by the Research Centres of Excellence programme supported by the National Research Foundation (NRF) Singapore and the Ministry of Education, Singapore. AB would like to acknowledge the National Research Foundation for financial support through grant number NRF2020-NRF-ISF004-3538.
\end{acknowledgments}

\section*{Data Availability}
The data that support the findings of this study are available from the corresponding author upon reasonable request.

\section*{REFERENCES}
\bibliography{library}

\end{document}